\documentclass{JHEP3}
\usepackage{amsmath}
\usepackage{amssymb}
\usepackage{graphicx}
\usepackage{cite}
\usepackage{tabularx}
\title{The deconfining phase transition in full QCD with two dynamical flavors}

\author{Paolo Cea \\ Dipartimento
Interateneo di Fisica, Universit\`a di Bari  and INFN - Sezione di Bari, \\
I-70126 Bari, Italy \\
E-mail:  \email{Paolo.Cea@ba.infn.it} }

\author{Leonardo Cosmai \\ INFN - Sezione di Bari, I-70126 Bari,
Italy\\
E-mail:  \email{Leonardo.Cosmai@ba.infn.it}}

\author{Massimo D'Elia\\ Dipartimento di Fisica, Universit\`a di Genova  and INFN - 
Sezione di Genova, \\
I-16146 Genova, Italy \\
E-mail:  \email{Massimo.Delia@ge.infn.it} }

\received{\today}

\abstract{
We investigate
the  deconfining phase transition in SU(3) pure gauge theory and in full QCD
with two flavors of staggered fermions. The phase transition
is detected by measuring the free energy in presence of an abelian
monopole background field.
In the pure gauge case our finite size scaling analysis 
is in agreement with the well known presence of a 
weak first order phase transition.
In the case of 2 flavors full QCD we find, using the standard pure gauge
and staggered fermion actions, that the phase
transition is consistent  with
weak first order, contrary to the expectation of a 
crossover for not too large quark masses and in agreement with 
results obtained by the Pisa group.}


\preprint{BARI-TH 2003/464 \\ GEF-TH/2003-15}

\begin{document}

\section{Introduction}
\label{Introduction}

Understanding QCD thermodynamics is one of the most intriguing issues of contemporary
physics. Indeed the study of QCD at high temperature 
(and density)~\cite{Satz:2000hm,Karsch:2001cy,Laermann:2003cv,Rischke:2003mt} is relevant
for both high energy physics (e.g. ultrarelativistic heavy ion collisions) and
astrophysics (compact stars). Moreover addressing QCD thermodynamics could shed light on the
problem of color confinement and chiral symmetry breaking.

To detect the deconfinement phase transition in pure gauge theories the expectation value
of the trace of the Polyakov loop is commonly used as an order parameter. 
In presence of dynamical fermions the  Polyakov loop ceases to be an order parameter since Z(N)
symmetry is no longer a symmetry of the action. Alternatively, in order to 
determine the finite temperature phase transition, one usually studies the chiral 
condensate, which however
is not related to confinement, but to chiral symmetry breaking, and in its 
turn is not a good order parameter at non-zero quark masses.

A mechanism for color confinement based on dual superconductivity of the 
QCD vacuum by abelian monopole condensation has been proposed 
in~\cite{tHooft:1976eps,Mandelstam:1974pi,Parisi:1975yh}.
A disorder parameter which is related to abelian monopole condensation 
in the dual superconductivity picture of confinement has been
developed by the Pisa group and consists in the vacuum expectation value of a 
magnetically charged operator, $\langle \mu \rangle$. In 
Refs.~\cite{DiGiacomo:1999fa,DiGiacomo:1999fb,Carmona:2001ja}
it has been shown that $\langle \mu \rangle$ is different from zero in
the confined phase of SU(2) and SU(3) pure gauge theories, that it goes to zero
at the deconfining phase transition, and that this is independent of the abelian
projection chosen to define the magnetic charge. The same has also 
been verified in the case of full QCD~\cite{Carmona:2002ty}.

In the present paper abelian monopole condensation is detected by looking at the 
free energy~\cite{Cea:2000zr,Cea:2001an}
in presence of an abelian monopole background field, which in turn is evaluated in terms of 
a gauge invariant lattice effective action.
This lattice effective action 
is defined at zero temperature by means of the lattice Schr\"odinger functional and at 
finite temperature by means of a thermal partition functional,
employed so far to study the vacuum dynamics of SU(3) lattice gauge theory 
at finite temperature in presence of a constant abelian background 
field~\cite{Cea:2002wx} or in presence of an abelian 
monopole background field~\cite{Cea:2000zr,Cea:2001an,Cea:2000rj}.

Since the free energy is related to the
vacuum dynamics and not, like the trace of the Polyakov loop, to a symmetry of the
gauge action which is washed out by the presence of dynamical fermions, 
we feel that it could be used
to detect the finite temperature phase transition also in full QCD, 
as well as the order parameter
$\langle \mu \rangle$ employed in Ref.~\cite{Carmona:2002ty}.

In the present paper we present a finite size scaling study of the free energy 
in presence of an abelian monopole background field both for SU(3) pure gauge 
and two-flavor QCD.
 
The plan of the paper is the following. 
In Sections~\ref{effaction} and~\ref{thermalpf}
we introduce the lattice effective action at zero temperature and at finite temperature.
In Section~\ref{Includingfermions} we modify the thermal partition functional for including
dynamical fermions.
In Section~\ref{SU3puregauge} we apply our method to the case of SU(3) pure gauge theory and,
by means of a finite size scaling analysis, we find that (as expected) the deconfining phase transition
is first order. In Section~\ref{QCDwith2df} we explore QCD thermodynamics with two
staggered dynamical fermions of equal masses using the standard staggered
fermion action. A finite size scaling analysis suggests 
that the deconfining phase transition is ``weak first order'', in agreement
with the indications presented in~\cite{Carmona:2002yg,Carmona:2002ty,Carmona:2003xs}.
In Section~\ref{Conclusions} we
sketch our conclusions. Finally in Appendix~\ref{AppendixU1}  we report 
our finite size scaling investigation of the U(1) lattice gauge theory.
A preliminary account of the results of the present paper appeared in 
Refs.~\cite{Cea:2002mr,Cea:2003un}.

\subsection{The lattice effective action: $T=0$}
\label{effaction}

In order to investigate vacuum structure of lattice gauge theories at zero temperature
a lattice gauge invariant effective action $\Gamma[\vec{A}^{\text{ext}}]$ for an external
background field $\vec{A}^{\text{ext}}$ was introduced in Refs.~\cite{Cea:1997ff,Cea:1999gn}.
It is defined as
\begin{equation}
\label{Gamma} 
\Gamma[\vec{A}^{\text{ext}}] = -\frac{1}{L_t} \ln
\left\{
\frac{{\mathcal{Z}}[\vec{A}^{\text{ext}}]}{{\mathcal{Z}}[0]}
\right\} \; .
\end{equation}
where $L_t$ is the lattice size in time direction.
$\vec{A}^{\text{ext}}(\vec{x})$  is the continuum gauge potential for the 
external static background field, the corresponding lattice links are
\begin{equation}
\label{links}
U^{\text{ext}}_k(\vec{x})= P \exp \int_0^1 dt \,\, iag A^{\text{ext}}_k(\vec{x}+ta\hat{k})
\end{equation}
($a$: lattice spacing, $g$: bare gauge coupling, $P$: path ordering operator).
${\mathcal{Z}}[\vec{A}^{\text{ext}}]$ is the lattice partition functional
\begin{equation}
\label{Zetalatt} 
{\mathcal{Z}}[\vec{A}^{\text{ext}}] = 
\int_{U_k(\vec{x},x_t=0)=U_k^{\text{ext}}(\vec{x})}
{\mathcal{D}}U \; e^{-S_W} \,, 
\end{equation}
$S_W$ is the Wilson action and 
the functional integration is performed over the lattice links, but constraining
the spatial links belonging to a given time slice (say $x_t=0$) to be
\begin{equation}
\label{coldwall}
U_k(\vec{x},x_t=0) = U^{\text{ext}}_k(\vec{x})
\,,\,\,\,\,\, (k=1,2,3) \,\,,
\end{equation}
$U^{\text{ext}}_k(\vec{x})$ being the lattice version (see Eq.~(\ref{links})) of the external 
continuum gauge field 
$\vec{A}^{\text{ext}}(x)=\vec{A}^{\text{ext}}_a(x) \lambda_a/2$  
(with $\lambda_a/2$ Gell-Mann matrices). The temporal links are not constrained.
In the case of a static background field which does not vanish at infinity we must also impose
that spatial links exiting from sites
belonging to the spatial boundaries (for each time slice $x_t \ne 0$) are fixed
according to Eq.~(\ref{coldwall}): in the continuum this last
condition amounts to the requirement that fluctuations over the
background field vanish at infinity.

The partition function defined in Eq.~(\ref{Zetalatt}) is also known as 
lattice Schr\"odinger functional~\cite{Luscher:1992an,Luscher:1995vs} and in the continuum 
corresponds to the Feynman kernel~\cite{Rossi:1980jf,Rossi:1980pg}.
Note that, at variance with the usual formulation of the lattice 
Schr\"odinger functional~\cite{Luscher:1992an,Luscher:1995vs} where a lattice
cylindrical geometry is adopted, our lattice has an hypertoroidal geometry so that 
$S_W$ in Eq.~(\ref{Zetalatt}) is allowed to be the standard Wilson action.

The lattice effective action $\Gamma[\vec{A}^{\text{ext}}]$ corresponds to the vacuum
energy, $E_0[\vec{A}^{\text{ext}}]$,
in presence of the background field with respect to the vacuum energy, $E_0[0]$,
with  $\vec{A}^{\text{ext}}=0$
\begin{equation}
\label{vacuumenergy}
\Gamma[\vec{A}^{\text{ext}}] \quad \longrightarrow \quad E_0[\vec{A}^{\text{ext}}]-E_0[0] \,.
\end{equation}
The relation above is true by letting the temporal lattice size $L_t \to \infty$,
on finite lattices this amounts to have $L_t$ sufficiently large 
to single out the ground state contribution to the energy.

Since the lattice effective action Eq.~(\ref{Gamma}) is given in terms of the 
lattice Schr\"odinger functional, which is invariant for time-independent gauge transformation of
the background field~\cite{Luscher:1992an,Luscher:1995vs}, it is as well gauge invariant.

\subsection{The thermal partition functional}
\label{thermalpf}
If we now consider the gauge theory at finite temperature $T=1/(a L_t)$ 
in presence of an external background field, the relevant quantity turns out to be
the free energy functional defined as
\begin{equation}
\label{freeenergy}
{\mathcal{F}}[\vec{A}^{\text{ext}}] = -\frac{1}{L_t} \ln
\left\{
\frac{{\mathcal{Z_T}}[\vec{A}^{\text{ext}}]}{{\mathcal{Z_T}}[0]}
\right\} \; .
\end{equation}
${\mathcal{Z_T}}[\vec{A}^{\text{ext}}]$ is the thermal partition functional~\cite{Gross:1981br}
in presence of the background field $\vec{A}^{\text{ext}}$, and is defined as
\begin{equation}
\label{ZetaTnew} 
\mathcal{Z}_T \left[ \vec{A}^{\text{ext}} \right]
= \int_{U_k(\vec{x},L_t)=U_k(\vec{x},0)=U^{\text{ext}}_k(\vec{x})}
\mathcal{D}U \, e^{-S_W}   \,,
\end{equation}
In Eq.~(\ref{ZetaTnew}), as in Eq.~(\ref{Zetalatt}), the spatial links belonging
to the time slice $x_t=0$ are constrained to the value of the external background field,
the temporal links are not constrained.
On a
lattice with finite spatial extension we also  usually impose that the
links at the spatial boundaries are fixed according to boundary
conditions Eq.~(\ref{coldwall}),  apart from the case in which the 
external background field vanishes at spatial infinity (as happens for 
the monopole field), where the choice
of periodic boundary conditions in the spatial
direction is equivalent to  Eq.~(\ref{coldwall}) in the thermodynamical limit.
If the
physical temperature is sent to zero, the thermal functional
Eq.~(\ref{ZetaTnew}) reduces to the zero-temperature Schr\"odinger
functional Eq.~(\ref{Zetalatt}).

The free energy functional Eq.~(\ref{freeenergy}) corresponds 
to the free energy, $F[\vec{A}^{\text{ext}}]$, in presence
of the external background field evaluated with respect 
to the free energy, $F[0]$, with 
$\vec{A}^{\text{ext}}=0$. When the physical temperature is sent to zero the free
energy  functional reduces to the vacuum energy functional Eq.~(\ref{Gamma}).

\subsection{Including fermions}
\label{Includingfermions}

When including dynamical fermions, the thermal partition functional
in presence of a static external background gauge field, Eq.~(\ref{ZetaTnew}),
becomes:
\begin{eqnarray}
\label{ZetaTfermions}
\mathcal{Z}_T \left[ \vec{A}^{\text{ext}} \right]  &=& 
\int_{U_k(L_t,\vec{x})=U_k(0,\vec{x})=U^{\text{ext}}_k(\vec{x})}
\mathcal{D}U \,  {\mathcal{D}} \psi  \, {\mathcal{D}} \bar{\psi} e^{-(S_W+S_F)} 
\nonumber \\ 
&=&  \int_{U_k(L_t,\vec{x})=U_k(0,\vec{x})=U^{\text{ext}}_k(\vec{x})}
\mathcal{D}U e^{-S_W} \, \det M \,,
\end{eqnarray}
where $S_W$ is the Wilson action, $S_F$ is the fermionic action and $M$ is 
the fermionic matrix.
Notice that the fermionic fields are not constrained and
the integration constraint is only relative to the gauge fields: 
this leads, as in the usual QCD partition function, to the appearance of 
the gauge invariant fermionic determinant after integration on the 
fermionic fields. 
As usual we impose on fermionic fields
periodic boundary conditions in the spatial directions and
antiperiodic boundary conditions in the temporal direction.

\subsection{The monopole free energy and the monopole condensation}
\label{monopolefreeenergy}
We use the free energy functional Eq.~(\ref{freeenergy}) to evaluate the free energy 
in presence of an abelian monopole background field (see Sect.~\ref{monopolebf}). 
If there is 
monopole condensation, the free energy with the monopole background field is the same
as the free energy without the monopole background field 
($\vec{A}^{\text{ext}}=0$). In other words it does not cost
energy to create a monopole. Therefore by evaluating the free energy 
by means of Eq.~(\ref{freeenergy}) we are able to detect abelian monopole condensation.

Note that, as discussed in Sect.~\ref{effaction}, our free energy functional is gauge invariant
for time-independent gauge transformations of the external background field. 
This implies that we 
do not need to do any gauge fixing to perform the abelian projection. 
Indeed, after choosing the abelian
direction, needed to define the abelian monopole field
through the abelian projection, due to gauge invariance of
Schr\"odinger functional for transformations of background field,
our results do not depend on the selected abelian direction,
which, actually, can be varied by a gauge transformation.
In this sense our definition is analogous to the definition given
in Ref.~\cite{Carmona:2001ja}, where the monopole field is defined
without gauge fixing.

To evaluate the monopole free energy by means of Eq.~(\ref{freeenergy})
we should compute the ratio of two partition functions. In order
to avoid the problem of dealing with  partition functions, we will compute
instead~\cite{Hasenfratz:1990tp,DelDebbio:1995sx}  the derivative of the free energy 
$F^\prime(\beta) = \partial {\mathcal{F}} /\partial \beta$ 
with respect to the gauge coupling $\beta$ ($\beta=6/g^2$). 
It is easy to see that $F^\prime(\beta)$
is given by the difference between the average plaquette
$<U_{\mu\nu}>$ obtained in turn from configurations with $\vec{A}^{\text{ext}}=0$
and $\vec{A}^{\text{ext}} \ne 0$
\begin{equation}
\label{fprime}
F^\prime(\beta) = \frac{\partial {\mathcal{F}}(\beta)}{\partial \beta} = 
V \left[ <U_{\mu\nu}>_{\vec{A}^{\text{ext}}=0} -
<U_{\mu\nu}>_{\vec{A}^{\text{ext}} \ne 0} \right] \,,
\end{equation}
where $V$ is the spatial volume. In Eq.~(\ref{fprime}) the dependence 
of the free energy  functional ${\mathcal{F}}$ on 
$\beta$ (at fixed external gauge potential $\vec{A}^{\text{ext}}$) has been made explicit.
Eventually, since ${\mathcal{F}}=0$ at $\beta=0$,
the free energy  can be obtained by numerical integration of $F^\prime(\beta)$ 
\begin{equation}
\label{trapezu1}
F(\beta)  = \int_0^\beta F^\prime(\beta^{\prime})
\,d\beta^{\prime}  \,.
\end{equation}
%

\section{SU(3) pure gauge}
\label{SU3puregauge}

\subsection{The monopole background field}
\label{monopolebf}

For SU(3) gauge theory the maximal abelian group is
U(1)$\times$U(1), therefore we may introduce two independent types
of abelian monopoles using respectively the Gell-Mann matrices 
$\lambda_3$ and $\lambda_8$ or their linear combinations.

In the following we shall consider the abelian monopole field related to 
the $\lambda_3$ diagonal generator.
In the continuum the abelian monopole field is given by
\begin{equation}
\label{monop3su2}
g \vec{b}^a({\vec{x}}) = \delta^{a,3} \frac{n_{\mathrm{mon}}}{2}
\frac{ \vec{x} \times \vec{n}}{|\vec{x}|(|\vec{x}| -
\vec{x}\cdot\vec{n})} \,,
\end{equation}
where $\vec{n}$ is the direction of the Dirac string and,
according to the Dirac quantization condition, $n_{\text{mon}}$ is
an integer. The lattice links corresponding to the abelian
monopole field Eq.~(\ref{monop3su2}) are (we choose $\vec{n}=\hat{x}_3$)
\begin{equation}
\label{t3linkssu3}
\begin{split}
U_{1,2}^{\text{ext}}(\vec{x}) & =
\begin{bmatrix}
e^{i \theta^{\text{mon}}_{1,2}(\vec{x})} & 0 & 0 \\ 0 &  e^{- i
\theta^{\text{mon}}_{1,2}(\vec{x})} & 0 \\ 0 & 0 & 1
\end{bmatrix}
\,  \\ U^{\text{ext}}_{3}(\vec{x}) & = {\mathbf 1} \,,
\end{split}
\end{equation}
with $\theta^{\text{mon}}_{1,2}(\vec{x})$ defined as
\begin{equation}
\label{thetat3su2}
\begin{split}
\theta^{\text{mon}}_1(\vec{x}) & = -\frac{n_{\text{mon}}}{4}
\frac{(x_2-X_2)}{|\vec{x}_{\text{mon}}|}
\frac{1}{|\vec{x}_{\text{mon}}| - (x_3-X_3)} \,, \\
\theta^{\text{mon}}_2(\vec{x}) & = +\frac{n_{\text{mon}}}{4}
\frac{(x_1-X_1)}{|\vec{x}_{\text{mon}}|}
\frac{1}{|\vec{x}_{\text{mon}}| - (x_3-X_3)} \,,
\end{split}
\end{equation}
where  $(X_1,X_2,X_3)$ are the monopole coordinates,
$\vec{x}_{\text{mon}} = (\vec{x} - \vec{X})$.
%
%
\FIGURE[ht]{\label{Fig1}
\includegraphics[width=0.85\textwidth,clip]{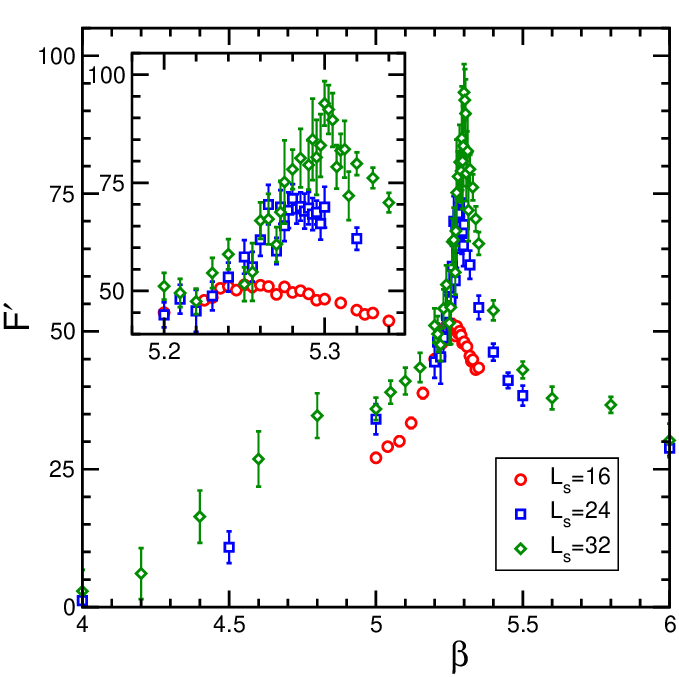}
\caption{
$F^{\prime}$ vs.  $\beta$  ($n_{\text{mon}}=10$)
for $L_s=16,24,32$ and $L_t=4$. Spatial ``fixed boundary conditions''
(for the significance of spatial ``fixed boundary conditions'' see Section~\ref{monopolebf}).}}
%
%
%
%
%
\FIGURE[ht]{ \label{Fig2}
\includegraphics[width=0.85\textwidth,clip]{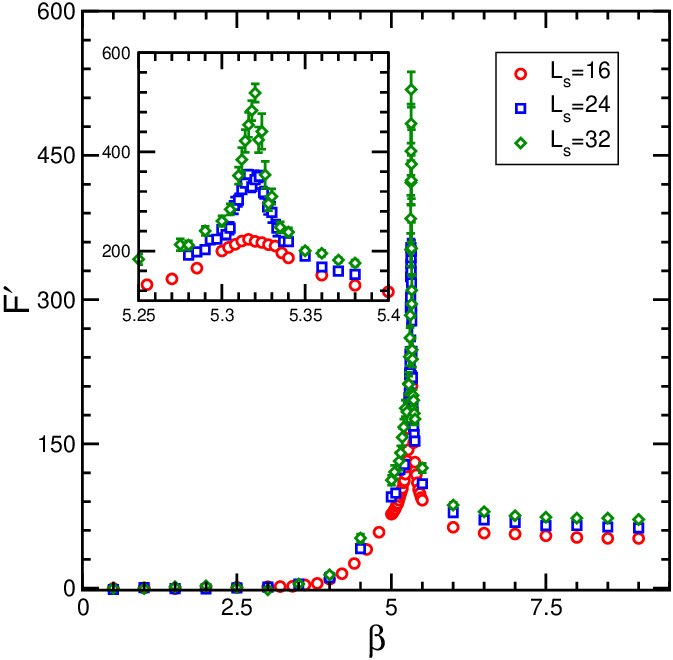}
\caption{
$F^{\prime}$ vs.  $\beta$  ($n_{\text{mon}}=10$)
for $L_s=16,24,32$ and $L_t=4$. Spatial ``periodic boundary conditions''
(for the significance of spatial ``periodic boundary conditions'' see Section~\ref{monopolebf}).} 
}

The monopole background field is introduced
by constraining (see Eq.~(\ref{t3linkssu3})) the spatial links exiting from the sites 
at the boundary of the time slice $x_t=0$. For what concern spatial links exiting from sites 
at the boundary of other time slices ($x_t \ne 0$) we consider two possibilities. 
In the first one we constrain these links according to Eq.~(\ref{t3linkssu3}) (in the
following we refer to this possibility as ``fixed boundary conditions''). 
In the second one we do not impose the constraint Eq.~(\ref{t3linkssu3}) on the above mentioned links 
(this possibility will be referred as ``periodic boundary conditions'').

We simulate pure SU(3) lattice gauge theory with Wilson action. The lattice geometry is
hypertoroidal. The path integral is constrained as given in Eq.~(\ref{t3linkssu3}). 
It is important to observe here that since the background field potential 
is not an integration variable (i.e. it does not correspond to a thermalized field)
it is not necessarily subject to periodic boundary conditions.

The simulations were performed on lattices of different spatial sizes taking
fixed the temporal extent ($L_t=4$). We consider lattices with spatial volumes
$16^3$, $24^3$, and  $32^3$. Simulations were performed in part on a APE100/QH1 
and in part on APEmille/crate in Bari. To upgrade SU(3) matrices we alternate
a Cabibbo-Marinari heat bath sweep with one (or more) overrelaxed sweep. The statistics
collected after 2,000 thermalization sweeps amounts at between 5,000 and 10,000 configurations
for each value of the gauge coupling $\beta$. The statistical analysis has been done by means of
jackknife resampling.

In Fig.~\ref{Fig1} we report our numerical results for
$F^\prime$ (Eq.~(\ref{fprime})) versus $\beta$ for three different
lattice sizes, $n_{\text{mon}}=10$, and spatial ``fixed boundary conditions''.
The data display a sharp peak that increases by increasing the lattice spatial
volume. The data reported in Fig.~\ref{Fig2} refer to spatial 
``periodic boundary conditions'' and display the same qualitative behavior, though with
an increased signal in the peak region. This is to be expected, since the effective volume is
larger than in the previous case with fixed spatial boundary conditions.

By inspecting Figure~\ref{Fig2} it is evident that $F^\prime(\beta)=0$ 
in a finite range of $\beta$, starting from $\beta=0$ and below the critical coupling,
signaled by a peak in $F^\prime(\beta)$. 
Therefore $F(\beta)=0$ in a finite range below the critical coupling,  above which 
the gauge system 
gets deconfined. The vanishing of the free energy implies abelian monopole condensation.

On the other hand, $F^\prime(\beta)$ becomes different from zero and increases with the lattice
spatial volume near the critical coupling, as expected in presence of a phase transition.
Moreover in the weak coupling regime $F^\prime(\beta)$ stays 
constant and almost independent from the spatial lattice volume\footnote{
We note a difference with the Pisa order parameter $\mu$:
in that case $\frac{d}{d \beta} \ln \langle \mu \rangle$ increases linearly 
with the spatial volume in the weak coupling limit, since
$\mu$ is a magnetically charged operator subject to  magnetic
charge superselection in the deconfined phase\cite{D'Elia:2003xn}.
}. 
This corresponds to the
classical monopole energy which depends linearly on $\beta$. It is clear that 
in the deconfined phase
it costs a finite amount of energy to create a monopole, and so,  there is not 
abelian monopole condensation.

In the following we will analyze the peak structure of $F^\prime(\beta)$ near the 
critical coupling and  show that the data display the right finite size scaling behavior
with scaling parameters compatible with a first order phase transition.

\subsection{Finite Size Scaling}
\label{fsspuregauge}

As a first step we determine the value of the critical coupling $\beta_c(L_s^{\text{eff}})$  
in correspondence
of each set of data (at different lattice sizes and different spatial boundary conditions).\\
We find that our data for $F^{\prime}(\beta,L_s^{\text{eff}})$ can be fitted according to 
\begin{equation}
\label{fitform1}
F^{\prime}(\beta,L_s^{\text{eff}}) 
= \frac{a(L_s^{\text{eff}})}{\left| \beta - \beta_c(L_s^{\text{eff}}) \right|^\alpha} \,, \quad
\beta < \beta_c(L_s^{\text{eff}}) \,.
\end{equation}
The quantity $L_s^{\text{eff}}$ is defined as 
\begin{equation}
\label{Leff}
L_s^{\text{eff}}   = 
\begin{cases}
L_s - 2 & \quad {\text{for spatial ``fixed b.c.'s''}}\,,  \\
L_s     & \quad {\text{for spatial ``periodic b.c.'s''}} \,.
\end{cases}
\end{equation}
The definition of $L_s^{\text{eff}}$ given in the previous Eq.~(\ref{Leff}) takes into account
the circumstance that for spatial ``fixed boundary conditions'' the effective spatial volumes
is reduced since the links exiting from the sites at the spatial boundaries of each time slice
($x_t \ne 0$) are constrained to the value given in Eq.~(\ref{t3linkssu3}).

Unlike the case of U(1) lattice gauge theory, discussed in~Appendix~\ref{AppendixU1},
where $\alpha \simeq 1$, we find that our data are best fitted with $\alpha \simeq 0.35$. 

In Table~\ref{Table1} 
we display $\beta_c(L_s^{\text{eff}})$ for both spatial periodic and fixed boundary condition.
\TABLE{
\begin{tabularx}{0.85\textwidth}{|XXX|}
\hline
\hline
\multicolumn{3}{|c|}{spatial ``fixed boundary conditions'' (see Section~\ref{monopolebf})} \\ \hline
$L_s$       &     $L_s^{\text{eff}}$    & $\beta_c$ \\ \hline
16          &     14                    &  5.4033 (959)    \\
24          &     22                    &  5.2976 (247)    \\
32          &     30                    &  5.3138 (108)    \\ \hline \hline
\multicolumn{3}{|c|}{spatial ``periodic boundary conditions'' (see Section~\ref{monopolebf})} \\ \hline
$L_s$       &     $L_s^{\text{eff}}$    & $\beta_c$ \\ \hline
16          &     16                    &  5.3296 (33)   \\
24          &     24                    &  5.3200 (56)   \\
32          &     32                    &  5.3199 (100)  \\ \hline \hline
\end{tabularx}
\caption{The value of $\beta_c$ in correspondence of different lattice sizes and 
different spatial boundary conditions, obtained by fitting Eq.~(\ref{fitform1}) to the
lattice data for $F^{\prime}(\beta,L_s^{\text{eff}})$ (see Eq.~(\ref{fprime})).}
\label{Table1}
}
From the various results for $\beta_c(L_s^{\text{eff}})$ we are able to get an estimate
of $\beta_c \equiv \beta_c(L_s^{\text{eff}}=\infty)$.
We try the following fit
\begin{equation}
\label{fitbetac}
\beta_c(L_s^{\text{eff}}) = \beta_c + d_1 (L_s^{\text{eff}})^{-1/\nu} \,.
\end{equation}
We find a rather good fit ($\chi^2/{\text{d.o.f.}}\simeq 0.8$) with $d_1 \simeq 0.6$
and
\begin{equation}
\label{nudafitbetac}
\beta_c = 5.3228 \pm 0.0024\,, \qquad \nu = 0.334 \pm 0.021 \,.
\end{equation}
The critical exponent $\nu$ is consistent with a first order phase
transition where $1/\nu = 3$.

%
\FIGURE[t]{ \label{Fig3}
\includegraphics[width=0.85\textwidth,clip]{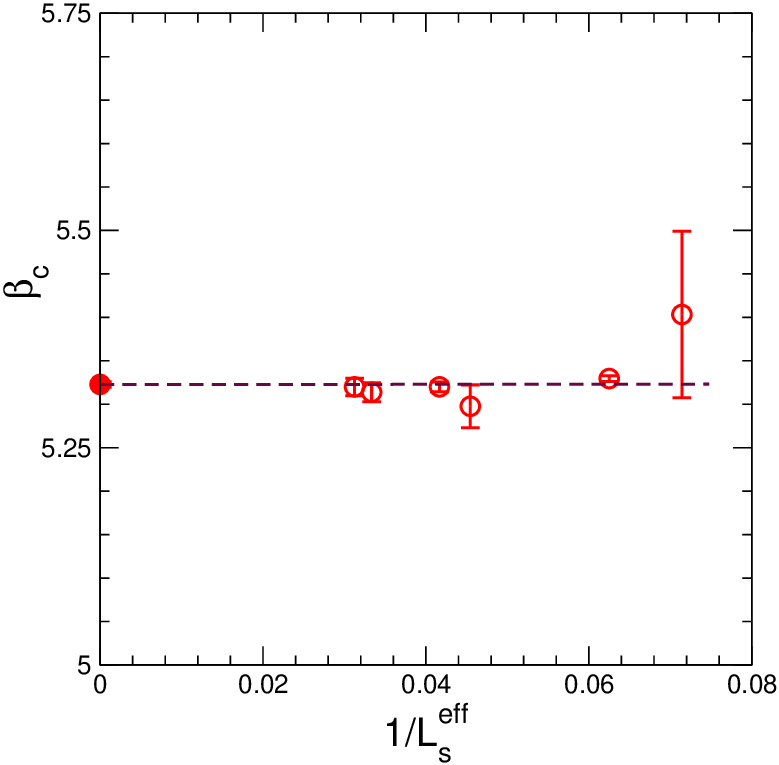}
\caption{The values of $\beta_c(L_s^{\text{eff}})$ (open circles), versus
$1/L_s^{\text{eff}}$, obtained 
at different values of $L_s^{\text{eff}}$ and different spatial boundary conditions (see Table~\ref{Table1}).
$L_s^{\text{eff}}$ is defined in Eq.~(\ref{Leff}).  The dotted line is  the fit
Eq.~(\ref{fitbetac}). The value $\beta_c(L_s^{\text{eff}}=\infty)$ is also plotted (full circle).} 
}

The behavior of $F^{\prime}(\beta,L_s^{\text{eff}})$ in the critical region can be investigated
by using finite size scaling techniques. Equations (\ref{fitform1}) and (\ref{fitbetac})
suggest the following scaling law
\begin{equation}
\label{Fprimefss}
F^{\prime}(\beta,L_s^{\text{eff}}) = \frac{a_1}{\left| \beta - \beta_c - d_1 (L_s^{\text{eff}})^{-1/\nu} \right|^\alpha} =
\frac{a_1 (L_s^{\text{eff}})^{\alpha/\nu}}{\left|  (L_s^{\text{eff}})^{1/\nu} (\beta - \beta_c) - d_1 \right|^\alpha} \,,
\end{equation}
so that $F^{\prime}(\beta,L_s^{\text{eff}}) / (L_s^{\text{eff}})^{\alpha/\nu}$ is a universal function of the
scaling variable 
\begin{equation}
\label{scalingvariable}
x =  (L_s^{\text{eff}})^{1/\nu} (\beta - \beta_c) \,.
\end{equation} 
Note that Eq.~(\ref{Fprimefss}) gives a sensible
result in the thermodynamical limit $L_s^{\text{eff}} \to \infty$
\begin{equation}
\label{fprimeLsinfinito}
F^{\prime}(\beta) = \frac{a_1}{|\beta - \beta_c|^\alpha}   \,, \quad \beta < \beta_c
\end{equation}
while $F^{\prime}(\beta=\beta_c,L_s^{\text{eff}})$ diverges like $(L_s^{\text{eff}})^{\alpha/\nu}$ when $L_s^{\text{eff}} \to \infty$.

Accordingly we fitted our lattice data with the scaling law
\begin{equation}
\label{scalinglaw}
F^{\prime}(\beta,L_s^{\text{eff}}) = \frac{a_1 (L_s^{\text{eff}})^{\gamma}}{\left|  (L_s^{\text{eff}})^{1/\nu} (\beta - \beta_c) - d_1 \right|^\alpha} \,,
\end{equation}
where we expect that $\gamma = \alpha/ \nu$.

\TABLE{
\begin{tabularx}{0.85\textwidth}{|XXXXXX|}
\hline
\hline
\multicolumn{6}{|c|}{spatial ``fixed boundary conditions'' (see Section~\ref{monopolebf})} \\ \hline
$a_1$       & $\gamma$   &  $\beta_c$   &    $\nu$   & $d_1$    & $\alpha$   \\ \hline
$12.199$    & $1.247$    &  $5.3251$    & $0.335$    & $0.6$    & $0.351$    \\
$\pm3.9004$ & $\pm0.089$ &  $\pm0.0110$ & $\pm0.026$ & constant & $\pm0.035$ \\ \hline \hline
\multicolumn{6}{|c|}{spatial ``periodic boundary conditions'' (see Section~\ref{monopolebf})} \\ \hline
$a_1$       & $\gamma$   &  $\beta_c$   &    $\nu$   & $d_1$    & $\alpha$   \\ \hline
$13.461$    & $1.510$    &  $5.3222$    & $0.340$    & $0.6$    & $0.347$    \\
$\pm1.337$ & $\pm0.555$ &  $\pm0.0013$ & $\pm0.020$ & constant & $\pm0.009$  \\ \hline \hline
\end{tabularx}
\caption{The values of the parameters obtained by fitting Eq.~(\ref{scalinglaw}) to the 
data for the derivative of the monopole free energy Eq.~(\ref{fprime}) 
on lattices with spatial volumes $16^3$, $24^3$, and $32^3$ and spatial ``fixed'' or ``periodic''
boundary conditions respectively.}
\label{Table2}
}

The output of the fits are reported in Table~\ref{Table2}. The parameter 
$d_1$ has been fixed at the value obtained with the fit Eq.~(\ref{fitbetac}).
We see clearly that $\alpha/\nu$ agrees with $\gamma$ within statistical errors,
as expected if we want a sensible result in the thermodynamical limit $L_s^{\text{eff}} \to \infty$
(see Eq.~(\ref{scalinglaw})).

In Figs.~\ref{Fig4},~\ref{Fig5} we plot $F^{\prime}(\beta,L_s^{\text{eff}})$ rescaled with the factor 
$1/(L_s^{\text{eff}})^{\gamma}$ 
versus the scaling variable $x$ defined in Eq.~(\ref{scalingvariable}), respectively
for spatial ``fixed boundary conditions'' and for spatial ``periodic boundary conditions''.
The full line is 
\begin{equation}
\label{scalingcurve}
\frac{F^{\prime}(\beta,L_s^{\text{eff}})}{(L_s^{\text{eff}})^{\gamma}} = 
\frac{a_1}{\left|  (L_s^{\text{eff}})^{1/\nu} (\beta - \beta_c) - d_1 \right|^\alpha} \,.
\end{equation}
We find that the scaling relation holds quite well for a very large range of $x$.
The quality of the scaling can be inferred looking at Figs.~\ref{Fig4} and~\ref{Fig5}.
Moreover looking at Table~\ref{Table2} we remarkably see that the critical 
parameters $\alpha$, $\beta_c$, $\nu$, and
$\gamma$ agree for both sets of data.

Finally in Fig.~\ref{Fig6} we display
$F^{\prime}(\beta,L_s^{\text{eff}}) / (L_s^{\text{eff}})^{\gamma}$ 
versus the scaling variable
$x$ for both periodic and fixed spatial boundary conditions. We see that
the numerical data nicely display the same scaling behavior. 

It is interesting to comment on the behavior of $\exp(-F(\beta)/T)$, which is the
analogous of disorder parameter of 
Refs.~\cite{DiGiacomo:1999fa,DiGiacomo:1999fb,Carmona:2001ja}, in the 
thermodynamical limit implied by Eq.~(\ref{fprimeLsinfinito}). Indeed we have that 
\begin{equation}
\label{Fdibeta}
\exp\left(-\frac{F(\beta)}{T}\right) = 
\exp \left( -\frac{1}{T} \int_{\beta_0}^\beta  F^{\prime}(\beta^\prime)\, d \beta^\prime  \right) \quad,
\quad \beta_0 < \beta < \beta_c  \,,
\end{equation}
while we already know that $\exp(-F(\beta)/T)=1$for $\beta < \beta_0$ irrespective of the 
lattice size (see Fig.~(\ref{Fig2})).
From Eq.~(\ref{fprimeLsinfinito}) we get
\begin{equation}
\label{Fdibetaint}
\exp\left(-\frac{F(\beta)}{T}\right) =  \exp\left[ - \frac{1}{T} \,\frac{a_1}{1-\alpha} 
\left( |\beta_0-\beta_c|^{1-\alpha}  - |\beta-\beta_c|^{1-\alpha}  \right)  \right]\,, \quad
\beta_0 < \beta < \beta_c  \,.
\end{equation}
So that $F(\beta)$ decreases when $\beta \to \beta_c$ if $0 < \alpha < 1$ tending to a finite value 
at $\beta = \beta_c$. On the other hand, for $\alpha=1$ it is easy 
to see that $\exp(-F(\beta)/T)$
decreases to zero as a power of $(\beta_c-\beta)$. Thus we conclude that for $\alpha < 1$
we have a discontinuous jump of $\exp(-F(\beta)/T)$ at $\beta_c$ 
and the strength of the discontinuity weakens when $\alpha \to 1$.
However, it must be stressed that the discontinuous jump of $\exp(-F(\beta)/T)$ 
is exceedingly small so that $\exp(-F(\beta)/T)$ decreases almost continuously 
toward zero when $\beta \to \beta_c$.
%
%
\FIGURE[!ht]{ \label{Fig4}
\includegraphics[width=0.85\textwidth,clip]{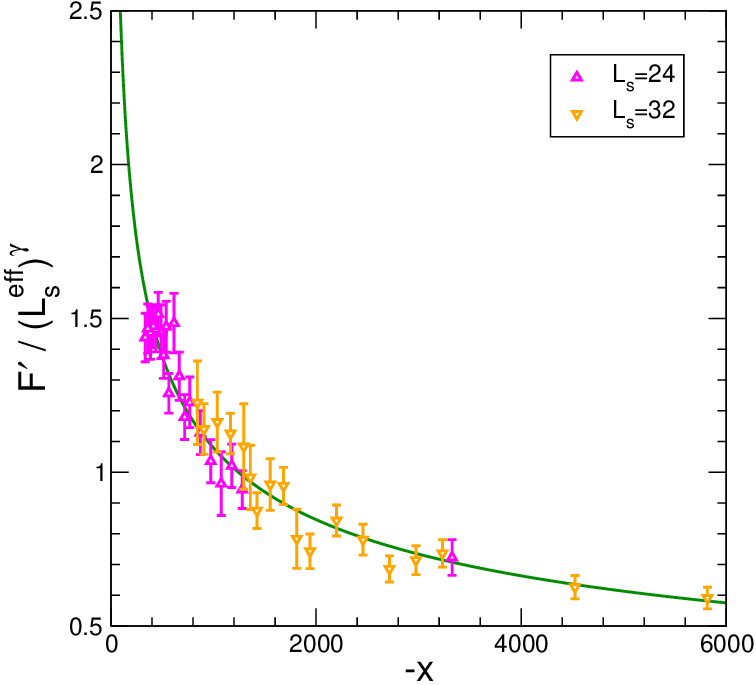}
\caption{The derivative of the monopole free energy with respect to the gauge coupling
$\beta$ (Eq.~(\ref{fprime})) rescaled with  $(L_s^{\text{eff}})^{\gamma}$ versus the
scaling variable $x$ (Eq.~(\ref{scalingvariable})). Data refer to simulations
with spatial ``fixed boundary conditions''. The full line is the scaling curve
Eq.~(\ref{scalingcurve}) with the parameters from Table~\ref{Table2}.}}
%
%
%
%
%
\FIGURE[!ht]{ \label{Fig5}
\includegraphics[width=0.85\textwidth,clip]{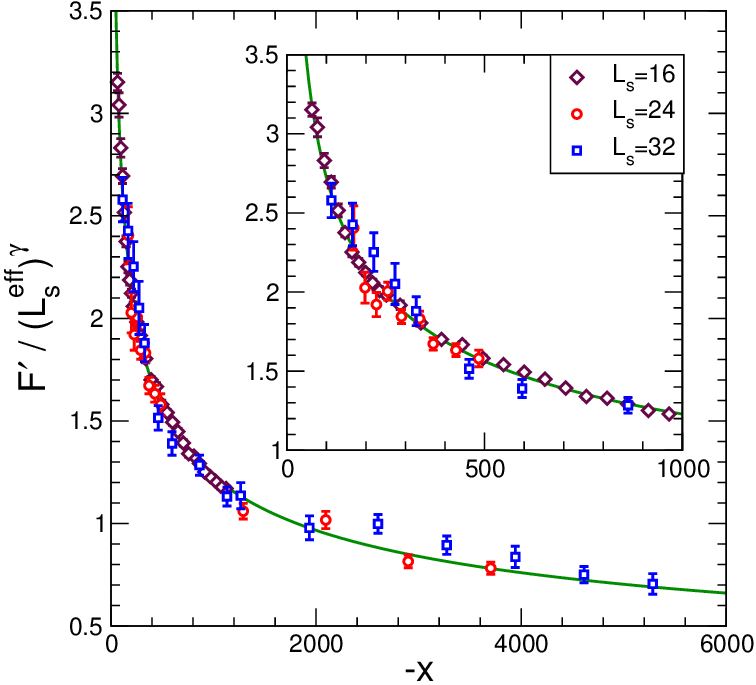}
\caption{The derivative of the monopole free energy with respect to the gauge coupling
$\beta$ (Eq.~(\ref{fprime})) rescaled with  $(L_s^{\text{eff}})^{\gamma}$ versus the
scaling variable $x$ (Eq.~(\ref{scalingvariable})). Data refer to simulations
with spatial ``periodic boundary conditions''. The full line is the scaling curve
Eq.~(\ref{scalingcurve}) with the parameters from Table~\ref{Table2}.}}
%
%
%
%
%
\FIGURE[!ht]{ \label{Fig6}
\includegraphics[width=0.85\textwidth,clip]{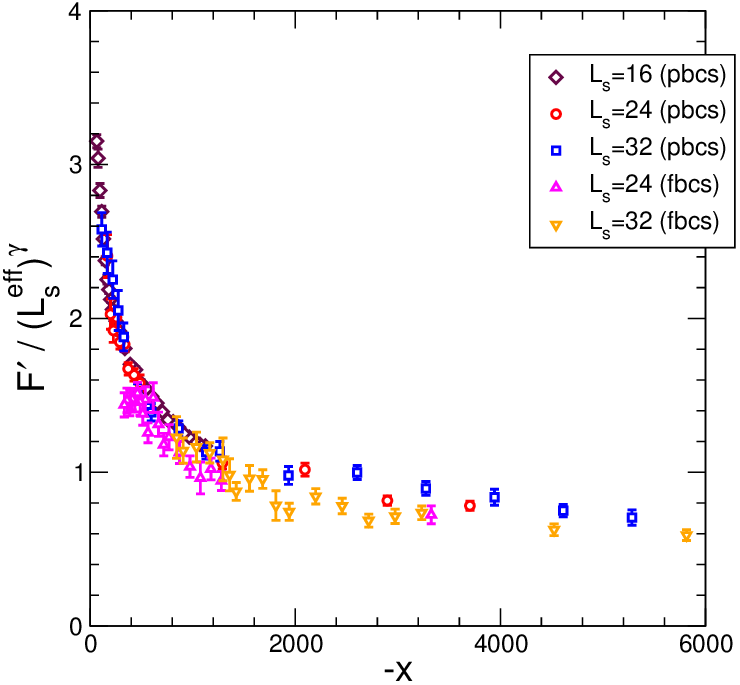}
\caption{We plot together data from Fig.~\ref{Fig4} (fbcs)  and Fig.~\ref{Fig5} (pbcs).
}}
%
%
%
%
%
%
%
%

\subsection{The deconfinement temperature}
\label{deconfinementtemperature}

Before ending the discussion of pure SU(3) gauge theory, we would like to remark 
that using the data for $F^{\prime}(\beta)$ we are able to get an estimate
of the continuum extrapolated critical temperature $T_c$ in units of
the square root of the string tension $\sqrt{\sigma}$. 
We will show here that our estimate is consistent,
even though with a quite large error, with the updated value in the literature.

In fact, by fitting Eq.~(\ref{fitform1}) to our data for $F^{\prime}(\beta)$
obtained on lattices  $32^3 \times L_t$ ($L_t=6,7,8$, with spatial
periodic boundary conditions), we are able to get 
an estimate of the critical coupling $\beta_c(L_t)$ corresponding to each $L_t$.
To fix a physical scale we consider the string tension $a \sqrt{\sigma}$
at each value of $\beta_c(L_t)$. 
The string tension is obtained on a symmetric lattice with the Wilson action.
To this purpose we use the data for the string tension as parameterized 
in Eq.(4.4) of Ref.~\cite{Edwards:1998xf}.

In Fig.~\ref{Fig7} our data for $T_c/\sqrt{\sigma}$ are displayed versus $a T_c$.
The continuum extrapolation using an ansatz quadratic in $aT_c$ gives the  estimate
$T_c/\sqrt{\sigma}=0.635\pm0.147$ to be confronted with the average estimate in the literature
$T_c/\sqrt{\sigma}=0.640\pm0.015$ (see Table 3 of Ref.~\cite{Teper:1998kw}).

Our conclusion is that our method allows to get an estimate of $T_c/\sqrt{\sigma}$
although the statistical uncertainty is quite large, mainly due to a 
large error in the evaluation of $\beta_c(L_t=6)$.
A better estimate of $\beta_c$ and then of 
$T_c/\sqrt{\sigma}$ could be achieved by means of the density spectral method.

%
%
%
%
%
\FIGURE[!ht]{ \label{Fig7}
\includegraphics[width=0.85\textwidth,clip]{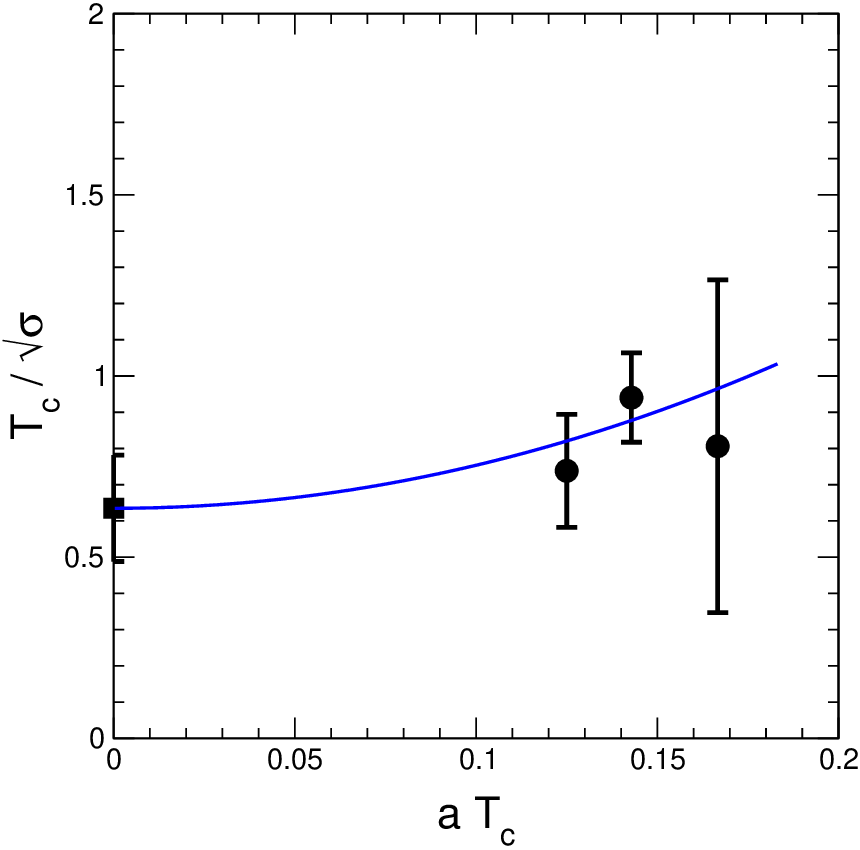}
\caption{Continuum extrapolation for $T_c/\sqrt{\sigma}$ (full circles). The extrapolated
value is $T_c/\sqrt{\sigma}=0.635\pm0.147$ (full square).}}
%
%
%
%

%
\section{QCD with two dynamical flavors}
\label{QCDwith2df}

We will account for results achieved from the study of QCD with two dynamical fermions using the standard staggered fermion action.
As discussed in the Introduction, we simulate the theory with  a ``cold'' time-slice (say $x_t=0$) 
where the spatial links are constrained to be a (lattice) abelian monopole background 
field (Eq.~(\ref{t3linkssu3})).  According to Section~\ref{monopolebf} the spatial 
links exiting from the sites belonging to the boundaries of the $x_t\ne0$ temporal slices 
are also constrained according to Eq.~(\ref{t3linkssu3}):
we have not considered the possibility of spatial periodic boundary
conditions in this case.

We compute the derivative of the monopole background field free energy 
with respect to the gauge coupling, as given in Eq.~(\ref{fprime}), 
where now the expectation value is evaluated with the full QCD action (details and 
numerical results in Sections~\ref{Numericalsimulations} and~\ref{Numericalresults}).
Our main goal is to try to use our data for $F^{\prime}(\beta)$ on different spatial 
volumes and different bare quark masses to infer the critical behavior of two flavors 
full QCD  near the deconfining transition.

\subsection{Numerical simulations}
\label{Numericalsimulations}

We used a slight modification of the standard HMC R-algorithm~\cite{Gottlieb:1987mq}
for two degenerate flavors of
staggered fermions with quark mass $m_q$: the links which are frozen
are not evolved during the molecular dynamics trajectory and the corresponding conjugate
momenta are set to zero.
We have collected about 2000
thermalized trajectories for each value of $\beta$ at $L_s=16,20$
and about 1000 thermalized trajectories for each value of $\beta$
at $L_s=32$. Each trajectory consists of $125$  molecular dynamics
steps and has total length  $1$. The computer simulations have
been performed on the APEmille crate.

\subsection{Numerical results}
\label{Numericalresults}

%
%
%
\FIGURE[!ht]{ \label{Fig8}
\includegraphics[width=0.85\textwidth,clip]{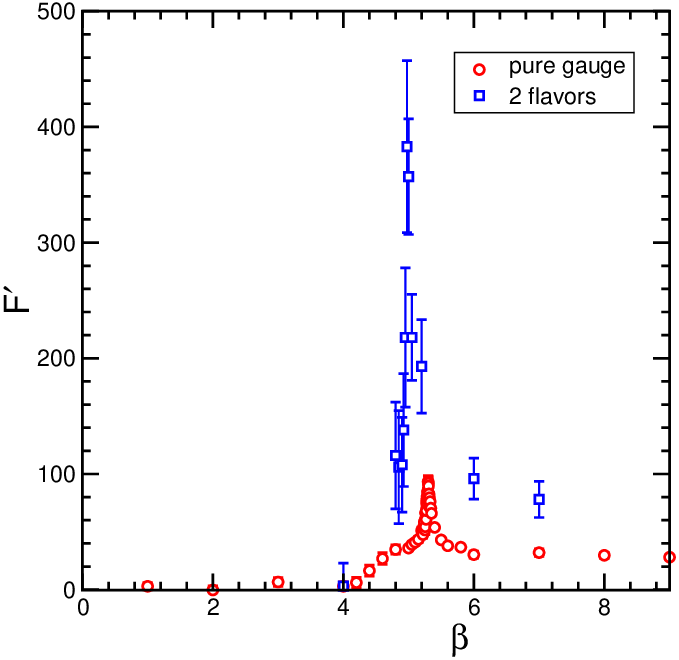}
\caption{
$F^{\prime}$ vs.  $\beta$  in the case of pure SU(3) gauge theory (open circles)
and full QCD with 2 dynamical flavors (open squares). Lattice size is $32^3\times4$.}}
In Fig.~\ref{Fig8} we compare $F^{\prime}(\beta)$ for two staggered degenerate flavors
with quark mass $m_q=0.075$ and in the quenched case. In the 2 flavors full QCD case the signal 
in the peak region gets enhanced with respect to the quenched case and the position
of the peak shifts to a smaller value of $\beta$.

%
%
\FIGURE[!ht]{ \label{Fig9}
\includegraphics[width=0.85\textwidth,clip]{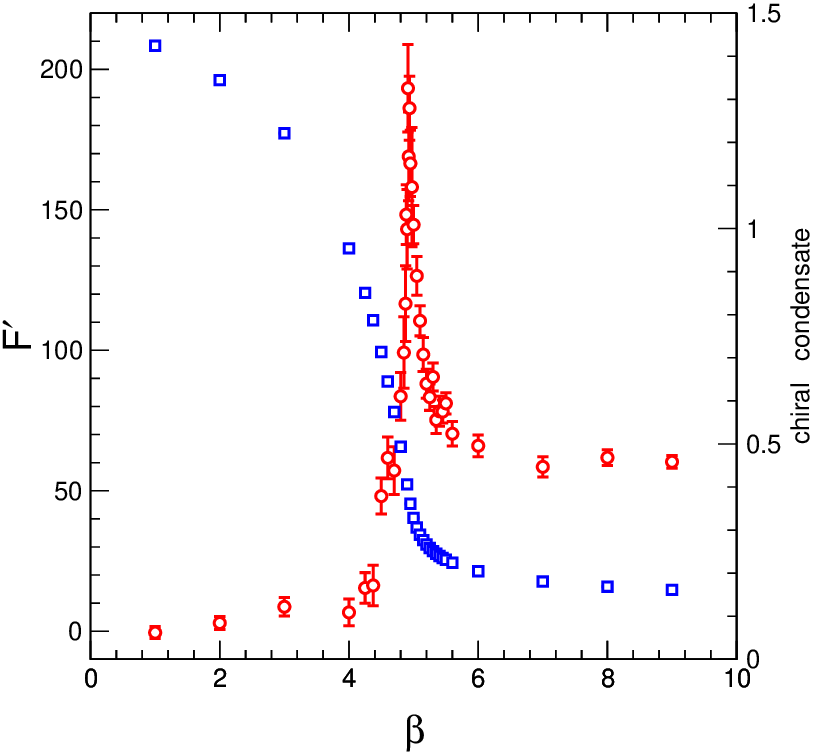}
\caption{
$F^{\prime}$  on a $16^3\times4$ lattice in the case of full QCD with 2 dynamical flavors, is displayed
together with chiral condensate.}}
In Fig.~\ref{Fig9} we display $F^{\prime}(\beta)$ in the case of 2 dynamical flavors
($m_q=0.075$, $16^3\times4$ lattice)
compared with the chiral condensate $<\psi \bar{\psi}>$.
Data displayed in Fig.~\ref{Fig9} suggest that the peak in $F^{\prime}(\beta)$
corresponds to the drop of the chiral condensate.
We varied the lattice size $L_s$ ($L_s=16,20,32$)  
and the staggered quark mass $m_q$ ($m_q=0.075,0.2676,0.5003$). 
At fixed $m_q$, as in the quenched
case, we find that the sharp peak increases with the lattice spatial volume. 
Moreover at fixed spatial volume the critical coupling depends on $m_q$.

\subsection{Finite Size Scaling}
\label{FSSfermions}
We would like to use the numerical data collected at different lattice size 
and staggered quark mass to perform a finite size scaling analysis.
According to Eq.~(\ref{Fprimefss}) we try the scaling law
\begin{equation}
\label{fssfermions}
F^{\prime}(\beta,L_s^{\text{eff}}) = 
\frac{a_1 (L_s^{\text{eff}})^{\gamma}}{\left|  (L_s^{\text{eff}})^{1/\nu} (\beta - \beta_c(m_q)) - d_1 \right|^\alpha} \,,
\end{equation}
where the critical coupling $\beta_c(m_q)$ depends on the quark mass $m_q$.
The dependence of the critical coupling $\beta_c(m_q)$ on the quark mass
is determined by the chiral critical point~\cite{Karsch:1994hm,Engels:2001bq,Karsch:2000kv}. 
In the thermodynamical limit, by known universality arguments the critical couplings will scale like
\begin{equation}
\label{betacmq}
\beta_c(m_q) = \beta_c(m_q=0) + c m_q^{1/\beta\delta} \,,
\end{equation}
where $1/\beta\delta$ is a combination of critical exponents which for the case of 2-flavors QCD are 
expected to be those of the three-dimensional O(4) symmetric spin models:
\begin{equation}
\label{betadelta}
\frac{1}{\beta \delta} \simeq 0.5415  \,.
\end{equation}
Inserting Eq.~(\ref{betacmq}) into Eq.~(\ref{fssfermions})  we are lead to the following scaling law
\begin{equation}
\label{scalingfermions}
F^{\prime}(\beta,L_s^{\text{eff}},m_q) = 
\frac{a_1 (L_s^{\text{eff}})^{\gamma}}{\left|  (L_s^{\text{eff}})^{1/\nu} (\beta - \beta_c(0) -c m_q^{\eta}) - d_1 \right|^\alpha} \,,
\end{equation}
where again $\gamma = \alpha/\nu$ assures a sensible thermodynamical limit.
Note that, to take care of finite volume effects, the exponent $\eta$ is expected to be:
\begin{equation}
\label{eta}
\eta = \frac{\nu_c}{\nu} \,, \quad \nu_c = \frac{\nu^{\prime}}{\beta \delta} \,,
\end{equation} 
where $\nu^{\prime}$, $\beta$, and  $\delta$ are the chiral critical exponents.

Indeed, Eqs.~(\ref{scalingfermions}) and~(\ref{eta}) assure that in the scaling region
\begin{equation}
\label{phifermions}
\frac{F^{\prime}(\beta,L_s^{\text{eff}},m_q)}{(L_s^{\text{eff}})^{\gamma}} = \Phi((L_s^{\text{eff}})^{1/\nu}(\beta - \beta_c(0)), (L_s^{\text{eff}})^{1/\nu_c} m_q) \,.
\end{equation}
In our case the relevant chiral critical exponents are those of the three-dimensional O(4) symmetric
spin models where~\cite{Engels:2001bq}
\begin{equation}
\label{nuprime}
\nu^{\prime} = 0.7423 \,, \quad \nu_c = 0.4019 \,.
\end{equation}
In Table~\ref{Table3} we report the results obtained by fitting 
Eq.~(\ref{scalingfermions}) to all our lattice data.
\TABLE{
\begin{tabularx}{1.0\textwidth}{|XXXXXXXX|}
\hline
\hline
\multicolumn{8}{|c|}{spatial ``fixed boundary conditions'' (see Section~\ref{monopolebf})} \\ \hline
$a_1$   & $\gamma$  &  $\beta_c(0)$     &   $c$       &  $\eta$    &  $\nu$     & $d_1$    & $\alpha$   \\ \hline
$79.4$    & $2.00$    &  $4.9933$       &   $0.54$    &  $1.10$    &  $0.31$    & $0.6$    & $0.728$    \\
$\pm76.6$ & $\pm0.47$ &  $\pm0.0138$    &   $\pm0.11$ &  $\pm0.19$ &  $\pm0.03$ & constant & $\pm0.078$  \\ \hline \hline 
\end{tabularx}
\caption{The values of the parameters obtained by fitting Eq.~(\ref{scalingfermions}) to the 
data for the derivative of the monopole free energy Eq.~(\ref{fprime})  in two-flavors full QCD
on lattices with spatial volumes $16^3$, $24^3$, and $32^3$ and $L_t=4$.}
\label{Table3}
}
\\
From Table~\ref{Table3} we can see that $\alpha/\nu=2.35\pm0.34$ consistent with $\gamma=2.00\pm0.47$
(see Section~\ref{fsspuregauge}). Concerning the parameter $\eta$ we find that it is
poorly determined by our data. If we constrain $\eta$ in our fit we get 
$\eta = 1.10\pm0.19$
which, together with $\nu=0.31\pm0.03$ leads to (see Eq.~(\ref{eta}))
$\nu_c=0.34\pm0.07$ consistent with the value reported in Eq.~(\ref{nuprime}).
However if we release the constraint on $\eta$ our data can also be fitted with smaller
values for $\eta$ without altering significantly the other parameters.
Moreover by confronting the exponent $\alpha$ in Table~\ref{Table3} with the corresponding 
value for the SU(3) pure gauge in Table~\ref{Table2} we conclude that our data for full 
QCD with two dynamical flavors are compatible with a first order phase transition ($\nu=0.31\pm0.03$) 
but, in the sense discussed at the end of Section~\ref{fsspuregauge}, this is weaker than in the quenched case. 
Our results are in agreement with the indications
for a first order phase transition in full QCD with 2 dynamical flavours (in the 
same range of quark masses) obtained in 
Refs.~\cite{Carmona:2002yg,Carmona:2002ty,Carmona:2003xs}.

%
%
%
\FIGURE[!ht]{ \label{Fig10}
\includegraphics[width=0.85\textwidth,clip]{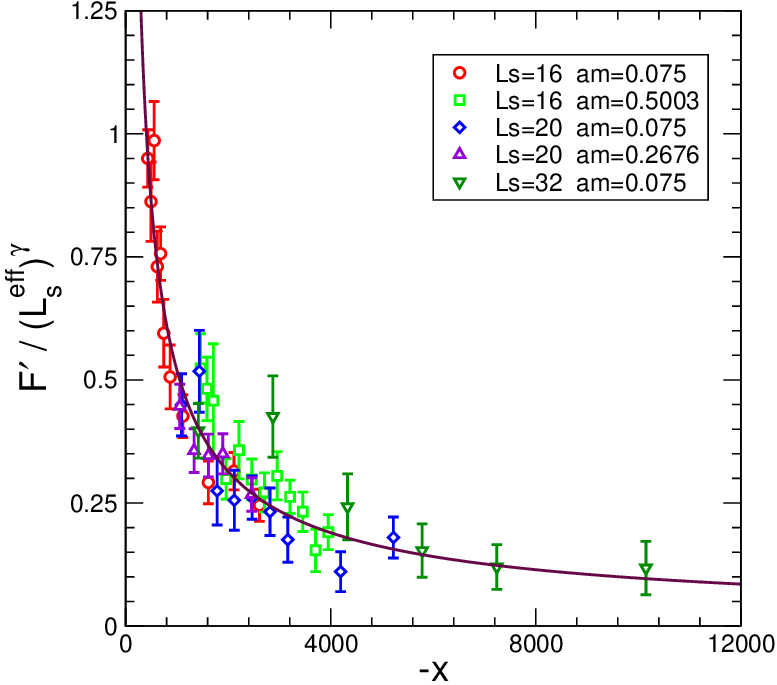}
\caption{$F^{\prime}(\beta,L_s^{\text{eff}},m_q)$ rescaled by the factor $(L_s^{\text{eff}})^\gamma$.
The values of $L_s^{\text{eff}}$ and $m_q=am$ are displayed in the legend.}}
%
%
%
%
%
\FIGURE[!ht]{ \label{Fig11}
\includegraphics[width=0.85\textwidth,clip]{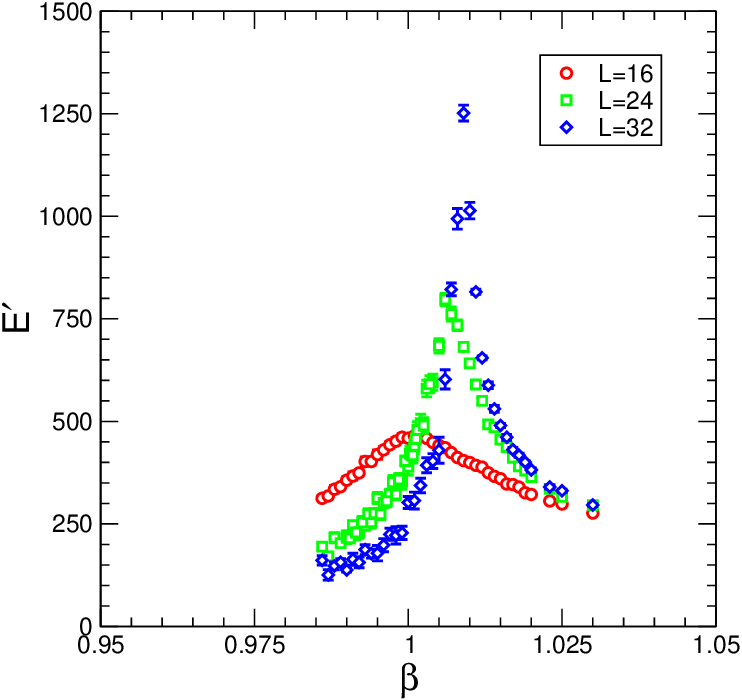}
\caption{
U(1) lattice gauge theory. The derivative with respect to $\beta$ of the monopole energy on lattices 
$L^4$ ($L=16,24,32$) and $n_{\text{mon}}=20$.}}
%
%
%
%
%
\FIGURE[!ht]{ \label{Fig12}
\includegraphics[width=0.85\textwidth,clip]{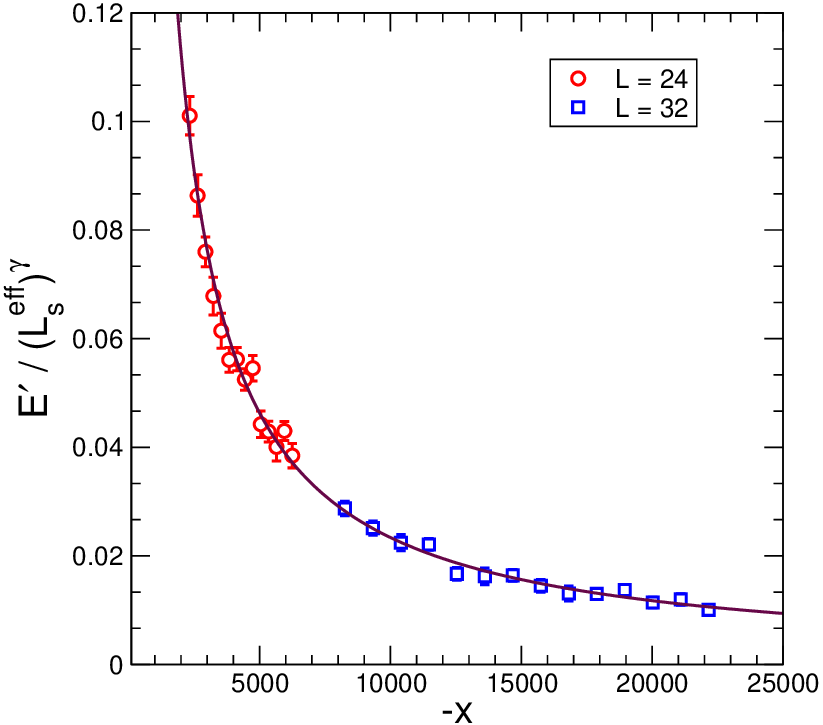}
\caption{U(1) lattice gauge theory. $E^{\prime}(\beta)$ rescaled by $(L_s^{\text{eff}})^\gamma$ 
on lattices $24^4$ and $32^4$. Solid line is the best fit according to the 
scaling law Eq.~(\ref{Fprimefss}).}}
%
%
%
%
%
%

%
\section{Conclusions}
\label{Conclusions}

Let us conclude by stressing the main results of this paper. We investigated 
the nature of deconfining phase transition in SU(3) pure gauge theory and in full QCD
with two flavors of staggered fermions. To locate the phase transition
we used the derivative of the monopole free energy with respect to the gauge coupling.
The monopole free energy is defined by means of a gauge invariant 
thermal partition functional in presence of the abelian monopole background field.
In the pure gauge case our finite size scaling analysis indicate
a weak first order phase transition. We get also an estimate of 
$T_c/\sqrt{\sigma}$ extrapolated to the continuum in good agreement 
with updated value in the literature. Moreover our method has been checked in U(1)
pure gauge theory giving results in agreement with previous investigations which
we present in Appendix A.

In the case of 2 flavors full QCD, we performed simulations by varying
spatial lattice sizes and quark masses. We find that deconfinement
transition in full QCD with 2 degenerate dynamical flavors is consistent  with
a weak first order phase transition, contrary to the expectation of a crossover 
for not too large quark masses, but in agreement with the indications
obtained with  
Refs.~\cite{Carmona:2002yg,Carmona:2002ty,Carmona:2003xs}.
Our results deserve further investigations that we
plan to do in the near future. 
In particular we would like to stress that we have  used the
standard pure gauge and staggered fermion action with $L_t = 4$, 
where finite lattice spacing scaling violations
could still be important, so that we plan to make use of an improved
action and/or of a larger value of $L_t$.
We plan to investigate the critical
region by means of the density spectral method. 
However, it is worthwhile to stress that, if this result will be confirmed 
the phase diagram of QCD with dynamical flavors should be reconsidered.

\acknowledgments{}
We greatly appreciate useful discussions with Adriano Di Giacomo.

\appendix

%
\section{U(1)}
\label{AppendixU1}
%
%
%

In this Appendix we take into account  pure gauge U(1) lattice 
theory at zero physical temperature, essentially for the purpose of 
showing that using our method the known results of compact U(1) are reproduced.
We consider U(1) lattice gauge theory in a monopole background field.
In the continuum the magnetic monopole field
with the Dirac string in the direction $\vec{n}$ is
\begin{equation}
\label{monopu1}
e \vec{b}^a({\vec{x}}) = \frac{n_{\mathrm{mon}}}{2}
\frac{ \vec{x} \times \vec{n}}{|\vec{x}|(|\vec{x}| -
\vec{x}\cdot\vec{n})} \,,
\end{equation}
where, according to the Dirac quantization condition,
$n_{\text{mon}}$ is an integer and $e$ is the electric charge
(magnetic charge = $n_{\text{mon}}/2e$). 
On the lattice the links belonging to the $x_t=0$ time-slice and
to the spatial boundaries of the $x_t\ne0$ time-slices are constrained
as (we choose $\vec{n}=\hat{x}_3$)
\begin{equation}
\label{monu1links}
\begin{split}
U^{\text{ext}}_{1,2}(\vec{x})  & = \cos [
\theta^{\text{mon}}_{1,2}(\vec{x}) ] + i  \sin [
\theta^{\text{mon}}_{1,2}(\vec{x}) ] \,, \\
U^{\text{ext}}_{3}(\vec{x}) & = {\mathbf 1} \,,
\end{split}
\end{equation}
with
\begin{equation}
\label{monu1theta}
\begin{split}
\theta^{\text{mon}}_1(\vec{x}) & = -\frac{n_{\text{mon}}}{2}
\frac{(x_2-X_2)}{|\vec{x}_{\text{mon}}|}
\frac{1}{|\vec{x}_{\text{mon}}| - (x_3-X_3)} \,, \\
\theta^{\text{mon}}_2(\vec{x}) & = +\frac{n_{\text{mon}}}{2}
\frac{(x_1-X_1)}{|\vec{x}_{\text{mon}}|}
\frac{1}{|\vec{x}_{\text{mon}}| - (x_3-X_3)} \,.
\end{split}
\end{equation}
In Equation~(\ref{monu1theta}) $(X_1,X_2,X_3)$ are the monopole
coordinates and $\vec{x}_{\text{mon}} = (\vec{x} - \vec{X})$. In
the numerical simulations we put the lattice Dirac monopole at the
center of the time slice $x_t=0$. To avoid the singularity due to
the Dirac string we locate the monopole between two neighboring
sites.

The abelian monopole condensation is detected by means of $E^{\prime}(\beta)$,
the derivative of the vacuum  energy (with respect to the gauge coupling) 
in presence of the monopole
background field and is given by (see Sect.~\ref{effaction}) 
\begin{equation}
\label{zerodisorder}
E = -\frac{1}{L_t} \frac{{\cal{Z}}[\vec{A}^{\text{ext}}]}{{\cal{Z}}[0]} \,
\end{equation}
where ${\cal{Z}}[\vec{A}^{\text{ext}}]$ is the lattice Schr\"odinger functional with 
$\vec{A}^{\text{ext}}$ the monopole background field and, according to the physical
interpretation of the effective action, we denote $\Gamma$ in Eq.~(\ref{Gamma}) with $E$.

Again, to avoid the problem of evaluating partition functions, 
we compute $E^\prime(\beta)$, the derivative with respect to $\beta$ of the monopole energy
$E(\beta)$, defined in terms of the lattice effective action Eq.~(\ref{Gamma}).
$E^\prime(\beta)$ is computed by the analogous of Eq.~(\ref{fprime}).

In Fig.~\ref{Fig11} the data for $E^\prime(\beta)$, with $n_{\text{mon}}=10$, on 
lattices $L^4$ ($L=16,24,32$) are displayed. We observe a sharp peak near the deconfining transition
($\beta \simeq 1.01$). To extract quantitative features from the lattice data we do 
a finite size scaling analysis.
To this purpose we employ the scaling law Eq.~(\ref{scalinglaw}) to fit the data for $E^\prime(\beta)$
on lattices $24^4$ and $32^4$. 
The best-fit parameters are reported in Table~\ref{Table4}.
In Fig.~\ref{Fig12} we report $E^{\prime}(\beta,L_s^{\text{eff}})$, rescaled by $1/(L_s^{\text{eff}})^\gamma$,
versus the scaling variable $(L_s^{\text{eff}})^{1/\nu}(\beta-\beta_c)$. As we can see the scaling law
Eq.~(\ref{scalinglaw}) holds very well for a wide range of the scaling variable $x$.
\TABLE{
\begin{tabularx}{0.85\textwidth}{|XXXXXX|}
\hline
\hline
\multicolumn{6}{|c|}{spatial ``fixed boundary conditions'' (see Section~\ref{monopolebf})} \\ \hline
$a_1$       & $\gamma$   &  $\beta_c$   & $\nu$      & $d_1$    & $\alpha$   \\ \hline
$285.2$     & $2.80$     &  $1.0107$    & $0.245$    & $200.0$  & $1.019$    \\
$\pm102.0$  & $\pm0.70$  &  $\pm0.0010$ & $\pm0.010$ & $102.8$  & $\pm0.036$ \\ \hline \hline
\end{tabularx}
\caption{The values of the parameters obtained by fitting Eq.~(\ref{Fprimefss}) to the 
data for the derivative of the monopole  energy Eq.~(\ref{fprime}) 
on lattices with  volumes $24^4$, and $32^4$.}
\label{Table4}
}
By inspecting Table~\ref{Table4} we can see that $\alpha/\nu=4.16\pm0.22$ which is consistent with 
$\gamma+1=3.8\pm0.7$. Indeed, according to Eq.~(\ref{zerodisorder}), to obtain a sensible result 
for $\beta < \beta_c$ in the thermodynamical limit,
we must have $\alpha/\nu = \gamma + 1$. Indeed, we see that this last relation is satisfied
within statistical uncertainty. Our results are in good quantitative agreement with 
previous finite size scaling study of the disorder parameter performed in Ref.~\cite{DiGiacomo:1997sm}.
Moreover our determination of the critical coupling $\beta_c$ is consistent with the most 
recent determination
reported in the literature~\cite{Arnold:2002jk}. 
For what concern the order of phase transition our data indicate a weak first order phase transition.
In fact the parameter $\nu$ is consistent with $1/d$ for $d=4$. 
Moreover  $\alpha\simeq1$ within errors 
indicating that the analogous of the disorder parameter 
of Refs.~\cite{DiGiacomo:1999fa,DiGiacomo:1999fb,Carmona:2001ja} (see the 
discussion at the end of Sect.~\ref{fsspuregauge})
goes to zero almost continuously as $\beta \to \beta_c$.
In this sense, we can say that the first order deconfinement transition in U(1)
is weaker than in SU(3).

\providecommand{\href}[2]{#2}\begingroup\raggedright\endgroup

\end{document}